\documentclass[twocolumn]{revtex4}

\usepackage{epsf}
\usepackage{graphicx}           

\begin{document}          


\title{Quantum data compression, quantum information generation, and the density-matrix renormalization group method}

\author{\"O.~Legeza and J. S{\'o}lyom}


\affiliation{Research Institute for Solid State Physics, H-1525 Budapest, P.\ O.\ Box 49, Hungary }

\date{\today}

\vskip -8pt
\begin{abstract}
We have studied quantum data compression for finite quantum systems where the site density matrices are not 
independent, i.e., the density matrix cannot be given as direct product of site density matrices and the 
von Neumann entropy is not equal to 
the sum of site entropies. Using the 
density-matrix renormalization group (DMRG)  
method 
for the 1-d Hubbard model, we have shown that a simple relationship exists between the entropy of the left or
right block and dimension of the Hilbert space of that block as well as of the superblock for any fixed accuracy.
The information loss during the RG procedure has been investigated and a 
more rigorous control of the relative error has been proposed based on Kholevo's theory.
Our results are also supported by the quantum chemistry version of DMRG applied to various molecules with system lengths up to 60 lattice sites.
A sum rule which relates site entropies and the total information generated by the renormalization 
procedure has also been given which serves as an alternative test of convergence of the DMRG method. 
\end{abstract}

\pacs{PACS number: 75.10.Jm}

\maketitle

\section{Introduction}

In the past few years non-local generalizations \cite{xiang,white3} of the density-matrix renormalization group (DMRG) method
\cite{white1,white2} have gained much attention since the method allows to study
interacting spin and electron systems in
momentum space \cite{xiang,erick1,legeza3} or
molecules within the context of quantum chemistry \cite{white3,white4,chan,mitrushenkov,legeza1,legeza2,legeza3}.

Recently the concepts of information theory have reappeared in the study of solid state physics 
and statistical physics problems
\cite{legeza3,vidal10,vidal11,korepin1,vidal12}.
On the one hand, it has been pointed out 
\cite{vidal10,vidal11} that the von Neumann entropy  of a finite system of length $N$, 
\begin{equation}
S(\rho^{(N)}) = -{\rm Tr}(\rho^{(N)} \ln \rho^{(N)})\,,
\label{eq:neumann}
\end{equation}
where $\rho^{(N)}$ is the density matrix,  
diverges logarithmically with the system size if the system is critical and the spectrum is gapless, while
it saturates as $N$ increases for non-critical, gapped models. 
On the other hand, in an independent work \cite{legeza3}, the von Neumann entropy has been used to improve the performance of
the momentum space and quantum chemistry versions of the DMRG method. 
It has been shown that an optimal ordering of states can be obtained from the
distribution of site von Neumann entropies of ``lattice sites''.
Quite recently, the study of entanglement has also led to a new procedure using periodic boundary condition \cite{verstraete}
that increases dramatically the performance of real space DMRG (RS-DMRG) method and   
to the development of time-dependent DMRG algorithm \cite{timedmrg1,timedmrg2}.

An important problem in quantum information theory is quantum data compression 
\cite{schumacher1,jozsa1,jozsa2,jozsa3}, i.e., how the dimension of the typical subspace should be chosen to achieve a prescribed fidelity.
The DMRG method is based on a similar reduction of the Hilbert space.
In the standard DMRG procedure the dimension, the number of states to be kept, is fixed before the calculation. In the 
dynamical block state selection  
(DBSS) approach \cite{legeza1} the truncation error is kept fixed and the dimension is allowed to vary. 
It has been shown \cite{legeza3} that the prescribed accuracy can be achieved with the least number of block states,
if the entropy-optimized ordering is used.   
In this paper, we study this problem from the point of view of quantum data compression and show that a simple
relationship exists between the von Neumann entropy of subsystem blocks and the size of Hilbert space 
of the blocks or alternatively the size of Hilbert space that one has to diagonalize in DMRG. 
We argue that the accessible information
\cite{kholevo1,fuchs1} of mixed-state ensembles  can be interpreted in the context of DMRG as the information loss due to the truncation 
procedure. Relying on this we propose a new method to improve the convergence of DMRG. 

The main purpose of this paper is to study the properties of DMRG from the point of view of
quantum information theory and to use the obtained results to optimize the method.
The setup of the paper is as follows. In Sec.~II we describe the theoretical background of
data compression, accessible information for mixed-state ensembles, and their relation to the DMRG method.
In Sec.~III these concepts and quantities are studied by applying the RS-DMRG 
to the half-filled 1-d Hubbard chain using open and periodic boundary conditions
and the non-local generalization of DMRG to various molecules.
In Sec.~IV the obtained results are interpreted as  
quantum information generation in the DMRG method.   
Our conclusions are presented in Sec.~V.

\section{Quantum Data Compression}

In his landmark work \cite{shannon}, Shannon has shown how much a message 
constructed from $N$ independent letters $(x_\alpha)$, where each letter occurs with a priori probability $p_\alpha$,  
can be compressed.
In classical information theory messages are   
classified into so-called typical and atypical sequences. If for any $\epsilon>0$
the sum of the probabilities of all typical sequences lies between $1-\epsilon$ and 1, then
for any $\delta>0$ and for large enough values of $N$ 
the number of typical sequences $n(\epsilon,\delta)$ lies between the bounds
$2^{N(S+\delta)}\geq n(\epsilon,\delta)\geq(1-\epsilon)2^{N(S-\delta)}$, where $S$ is the Shannon entropy, $S=-\sum_\alpha p_\alpha \log_2 p_\alpha$. 
Therefore, a block code of length $NS$ bits encodes all typical sequences
irrespective of how the atypical sequences are encoded and the probability of error will still be less than $\epsilon$.

In quantum information theory the letters are density matrices and  
one has to distinguish two cases, namely when the density matrices correspond to ensembles of $q$ pure states, 
$|\phi_\alpha\rangle$, or when they are formed from density matrices $\rho_\alpha$, with probability $p_\alpha$.

\subsection{Pure-state ensemble}

Considering the first case, that is a pure-state ensemble the   
density matrix of a message consisting of $N$ letters is
$\rho^{(N)}=\rho\otimes\rho\otimes ... \otimes\rho$, where   
$\rho=\sum_\alpha p_\alpha |\phi_\alpha\rangle\langle\phi_\alpha|$, and the
von Neumann entropy of the message is simply related to the entropy of the ensemble,  
$S(\rho^{(N)})=N S(\rho)$. 
The optimal code that compresses the  
Hilbert space of the entire message, 
$\Lambda^{(N)} = \Lambda\otimes {\Lambda}\otimes \cdots \Lambda$
to a smaller Hilbert space without compromising
the fidelity of the message for $N\rightarrow\infty$ 
has been obtained by B.~Schumacher \cite{schumacher1}.
Analogously to classical information theory $\Lambda^{(N)}$ is divided into so-called typical 
(${\Lambda}_{\rm typ}^{(N)}$) and atypical (${\Lambda}_{\rm atyp}^{(N)}$)
subspaces by applying 
a projection $\Pi_{\rm typ}$ and $\Pi_{\rm atyp}$. If for any $\epsilon>0$ almost all weight of the ensemble lies within $\Lambda^{(N)}_{\rm typ}$ and
$Tr\Pi_{\rm typ}\rho^{(N)}\Pi_{\rm typ} > 1-\epsilon$ while for the atypical subspace 
$Tr\Pi_{\rm atyp}\rho^{(N)}\Pi_{\rm atyp} < \epsilon$, then for any $\delta>0$ and sufficiently large enough $N$ 
the eigenvalues $\omega_{\rm typ}$ of  $\rho^{(N)}$ belonging to typical eigenstates 
fall within a narrow range:
\begin{equation}
e^{-N[S(\rho)+\delta ]} < \omega_n < e^{-N[ S(\rho)-\delta]}\,.
\label{eq:qdc01}
\end{equation}
Therefore, the number of dimensions of the typical subspace lies between the bounds 
\begin{equation}
(1-\epsilon)e^{N[S(\rho)-\delta ]} \leq \dim{\Lambda}_{\rm typ}^{(N)} \leq e^{N[ S(\rho)+\delta ]}\,.
\label{eq:qdc02}
\end{equation}
This means that the von Neumann entropy is the number of qubits of quantum information carried per letter of the message and
unless $\rho=\frac{1}{q}{\openone}$,  a compression is always possible. We have to mention that Eqs.~(\ref{eq:qdc01})-(\ref{eq:qdc02})
have been reformulated in order to be consistent with Eq.~(\ref{eq:neumann}).

\subsection{Mixed-states ensemble}

The result is less simple if the letters are chosen from density matrices of mixed states. 
In this case only an upper \cite{kholevo1} or lower bound \cite{jozsa4} can be
derived for the accessible information.
Therefore, if the source of
information is constructed from messages represented by $\rho_a$ states and a priori probability $p_a$, then
the mutual information between the sender's and receiver's measurement is bounded by
\begin{equation}
I\leq S(\rho)-\sum_a p_a S(\rho_a)\,,
\label{eq:holevo}
\end{equation}
where $\rho=\sum_a p_a \rho_a$ and
$S$ is the von Neumann entropy given by Eq.~(\ref{eq:neumann}). 
If all signal states $\rho_a$ are pure states, 
the upper bound on the accessible information reduces to $I\leq S(\rho)$.

The lower bound on accessible information
also depends not only on the entire density matrix but also on the
particular way $\rho$ is realized as an ensemble of mixed states. 
As it has been given by Jozsa {\sl et al.} \cite{jozsa4}, 
\begin{equation}
I \geq Q(\rho) - \sum_\alpha p_\alpha Q(\rho_\alpha)\,,
\end{equation}
where the subentropy $Q$ is defined as
\begin{equation}
Q = -\sum_{\alpha=1}^M \left ( \prod_{\alpha\neq \beta} \frac{\omega_\alpha}{\omega_\alpha - \omega_\beta} \right ) \omega_\alpha \ln \omega_\alpha \,.
\end{equation}
If two or more of the eigenvalues are equal, $Q$ remains finite if one takes the limit $\omega_\alpha\rightarrow\omega_\beta$.

\subsection{Relationship to DMRG}

For lattice models, studied in solid state physics problems, on the other hand, 
the message coded into the wave function of the system has different features. The
site density matrices are in general not independent, thus  
$\rho^{(N)}\neq\rho\otimes\rho\otimes ... \otimes\rho$ and $S(\rho^{(N)}) \neq N S(\rho)$. In this paper,
this situation is studied using DMRG. 

\begin{figure}
\includegraphics[scale=0.35]{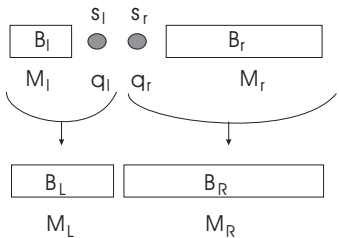}
\vskip .4cm
\caption{Schematic plot of the system and environment block of DMRG. $B_l$ and $B_r$ denote the left and right block
of length $l$ and $r$, and of dimension $M_l$ and $M_r$, respectively, $\bullet$ stands for the
intermediate sites ($s_l$ and $s_r$) with $q_l$ and $q_r$ degrees of freedom. The blocks $B_L = B_l\bullet, B_R = \bullet B_r$ have dimension
$M_L$ and $M_R$, respectively.}
\label{fig:dmrg}
\end{figure}

From the point of view of information theory DMRG is a numerical tool to select the typical subspace   
on which the target state ($\Psi_{\rm TG}$) of a Hamiltonian of a finite system of length $N$ can be represented. 
The Hilbert space of the system (called superblock Hilbert space) is defined on a bi-partite system,   
$\Lambda_{\rm typ} = \Lambda^{(L)}_{\rm typ}\otimes\Lambda^{(R)}_{\rm typ}$ where $\Lambda^{(L)}_{\rm typ}$ and $\Lambda^{(R)}_{\rm typ}$ 
are the typical Hilbert spaces of the left and right blocks, $B_L$ and $B_R$, which themselves are built from subblocks, $B_l$ and $B_r$,
with one extra site. The indices $l$ and $r$ denote at the same time the number of sites in the subblocks.
The schematic plot of the DMRG configuration is shown in Fig.~\ref{fig:dmrg}.

Although the target state is usually a pure state, the left and right blocks are in a mixed state 
described by the density matrices $\rho_L={\rm Tr}_R \rho$, and $\rho_R = {\rm Tr}_L \rho$, respectively. Analogously one can define
the density matrices of the $B_l$ and $B_r$ subblocks, 
$\rho_l={\rm Tr}_{s_l} {\rm Tr}_R \rho$, $\rho_r = {\rm Tr}_L {\rm Tr}_{s_r} \rho$,
as well as that of the intermediate sites, given as
$\rho_{s_l}={\rm Tr}_l {\rm Tr}_R \rho$, $\rho_{s_r} = {\rm Tr}_L {\rm Tr}_r \rho$.
It follows from 
singular value decomposition theorem that  
for a pure target state
and any choice of the length of the left subblock, $l$, the entropy of the left block 
of length $l+1$, $S_L(l+1)$ is identical to that of the right block, $S_R(r+1)$, where $l+r+2=N$.
$S\equiv S_L=S_R$    
is also related to the mutual information of the blocks, $I=-2S$ 
since $S_L(l+1) + S_R(r+1) + I = S(N)$, where $S(N)$ is the entropy of the full system,
which vanishes if the system is in a pure state.  
$I$ quantifies the correlation between the two subsystems 
and vanishes if and only if the two subsystems are completely uncorrelated, i.e., when $\rho = \rho_L\otimes\rho_R$.
$S$ also measures the entanglement \cite{entang1} 
and the amount of quantum information exchange between the blocks \cite{entang2}.

The wave function of the finite lattice of length $N$ is built up using the {infinite lattice} method followed by a systematic application of the sweeping 
procedure of the {finite lattice} method. The typical subspaces $\Lambda^{(L)}_{\rm typ}$ and $\Lambda^{(R)}_{\rm typ}$ are obtained  
by selecting that $M$ renormalized states out of the $M\times l$ block states ($|\phi_\alpha^{(L)}\rangle,|\phi_\alpha^{(R)}\rangle$), which have the 
largest eigenvalues, $\omega_\alpha^{(L)}$ and $\omega_\alpha^{(R)}$   
of the density matrices $\rho_L$ and $\rho_R$, respectively.
The $\Pi_{\rm typ}^L$ and $\Pi_{\rm typ}^R$ projection operators are formed from the $M$ chosen eigenstates of $\rho_L$ and $\rho_R$.  
The accuracy of the
truncation procedure is given by the so-called truncation error, 
$TRE = 1-\sum_{\alpha=1}^M \omega_\alpha$, where $\omega_\alpha \equiv \omega^{(L)}_\alpha$ or $\omega^{(R)}_\alpha$, depending on the direction of the sweep. 
Therefore, due to the truncation of basis states $\epsilon$ is always greater than zero.

Another main difference between conventional information theory and DMRG is that 
while in the former case the letter states have an a priori assigned probability density \cite{hausladen,schumacher2001}, in the latter case the probability of a configuration
depends on the target state. Therefore, the subset of states which define the code words is determined by the physical properties of the target state.
When DMRG is applied to find the typical subspace of a target state, i.e., the typical code words, this subspace is reduced further.
In addition, during the RG procedure
the original basis states are also transformed into new renormalized states, thus code words change at each iteration step.
In this respect, in DMRG the information of one block 
is coded into the basis function of the renormalized blocks 
rather than into a quantum channel of qubits.

In the standard DMRG procedure the number of states of the left and right blocks, $M$, is fixed in advance of the calculation. 
The superblock Hilbert space, $\Lambda_{SB}$, on which the target state is determined is not simply
$\Lambda^{(L)}_{\rm typ} \otimes \Lambda^{(R)}_{\rm typ}$, since only states with prescribed quantum numbers have to be considered.
Usually $M$ is increased systematically during the finite lattice method,  
thus the accuracy depends on $M$ in an uncontrolled way.
In contrast to this in the DBSS approach \cite{legeza1} the truncation error 
is kept fixed and the number of retained states is varied dynamically, 
which allows to control the accuracy more precisely. A threshold value on the minimum number of block states $M_{min}$ 
has also been introduced in order to avoid higher lying local minima.

\section{Numerical results}

The most general fermionic Hamiltonian can be given as
\begin{equation}
{\cal H} = \sum_{ij\sigma} T_{ij} c^\dagger_{i\sigma}c_{j\sigma} +
           \sum_{ijkl\sigma\sigma^\prime} {V_{ijkl}
c^\dagger_{i\sigma}c^\dagger_{j\sigma^{\prime}}c_{k\sigma^{\prime}}c_{l\sigma}},
\label{eq:ham}
\end{equation}
where $T_{ij}$ denotes the matrix elements of the one-particle
Hamiltonian and $V_{ijkl}$ stands for the matrix elements of the electron
interaction operator. Depending on the structure of $T_{ij}$ and $V_{ijkl}$ this
Hamiltonian can describe fermionic models in real space with open or periodic boundary condition,
a molecule or a usual fermionic model in momentum space.

A special case of Eq.~(\ref{eq:ham}) is 
the 1-d Hubbard model, 
\begin{equation}
{\cal H} = \sum_{i\sigma} t ( c_{i\sigma}^\dagger c^{\phantom\dagger}_{i+1\sigma} + c_{i+1\sigma}^\dagger c^{\phantom\dagger}_{i\sigma} ) + 
U \sum_i c_{i\uparrow}^\dagger c_{i\uparrow}^{\phantom\dagger} c^\dagger_{i\downarrow} c_{i\downarrow}^{\phantom\dagger}\,. 
\label{eq:hubrs}
\end{equation}
In this case the $|\phi_\alpha\rangle$ states are  
$|0\rangle$, $|\!\downarrow\rangle$, $|\!\uparrow\rangle$, $|\!\downarrow\uparrow\rangle$.
For homogeneous lattice models with periodic boundary condition each lattice site carries the same information ($S_i$). 
For $U=0$, $S_{i}=4\ln4=1.38629$ while for $U\rightarrow\infty$ the model is equivalent to the spin$-1/2$ Heisenberg model and
only the $|\!\downarrow\rangle$ and $|\!\uparrow\rangle$ states have finite weight, thus $S_{i}=2\ln 2= 0.69314$.
In the numerical work we have used the real space version of DMRG with open (OBC) and periodic (PBC) boundary conditions.

\subsection{The standard DMRG procedure}

\begin{figure}
\includegraphics[scale=0.38]{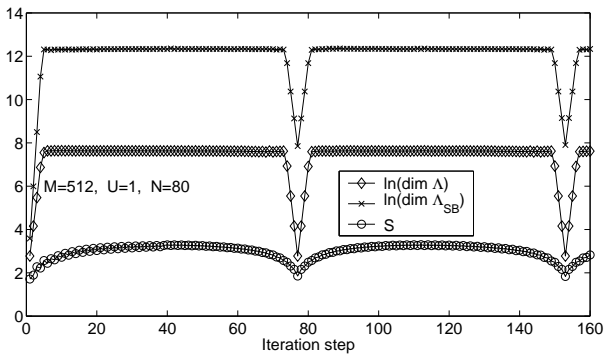}
\caption{Block entropy, $S$, and $\ln(\dim\Lambda)$, $\ln(\dim\Lambda_{SB})$ 
as a function of iteration steps for the half-filled 1-d Hubbard model with $N=80$ for $U=1$ 
with fixed number of block states ($M=512$) using PBC. For iteration step less than $39$ the infinite lattice method has been used. }
\label{fig:rsdmrg}
\end{figure}

First we have investigated the block entropy in the standard RS-DMRG method when calculating the ground state energy of the half-filled 1-d Hubbard model 
with $t=1$ for a chain with $N=80$ sites, for $U=1,10,100$ using a fixed number of block states, namely $M=256,512,1024$.  
In Fig.~\ref{fig:rsdmrg} we have plotted the block entropy  
and the logarithm of the dimension of $\Lambda^{}$ and the Hilbert space of the superblock, $\ln(\dim \Lambda),\ln(\dim \Lambda_{SB})$, respectively, 
as a function of iteration steps for $U=1$ and $M=512$.  
It is clear from the figure that for the {\em infinite lattice} method (for iteration steps less than 39) 
the block entropy, $S$, increases with the block length.
According to Vidal {\sl et al.} \cite{vidal10,vidal11} a logarithmic divergence is expected if no truncation had been applied during the RG procedure. 
Deviations occur for two reasons. First the truncation procedure has been applied from the 5th iteration step. More importantly, in DMRG the chains
are built up systematically, and for chains of length $N$ the finite lattice method has to be used after $N/2-1$ initialization steps. At this point the entropy
of the blocks starts to decrease. In the successive sweeps the entropy reaches its maximum when the sizes of the two blocks become equal.
It is worth mentioning that $S$ became almost totally symmetric after the second half sweep already. 
Since $S$ is related to the mutual information of the blocks 
this result is in agreement with the long known fact that 
the best accuracy of RS-DMRG is always obtained for the symmetric 
block configurations.  
For larger $U$ values the entanglement between the blocks is reduced as indicated by the decreasing value of $S$ and $S_i$, while $\ln (\dim \Lambda)$ 
remained almost the same, and the relative error of the calculations improved significantly. 
This is the first sign that the entropy, the dimension of the Hilbert space and the accuracy is related to each other in DMRG. 

\subsection{DBSS approach}

\begin{figure}
\includegraphics[scale=0.38]{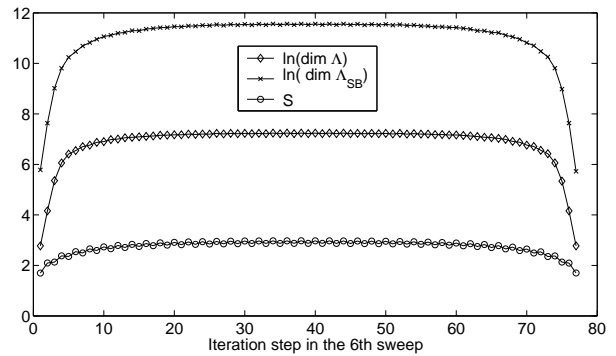}
\caption{Same as Fig.~2 but using DBSS approach for $TRE_{max}=10^{-4}$ and $M_{min}=16$.}  
\label{fig:qdc_dbss_u1}
\end{figure}

\begin{figure}
\includegraphics[scale=0.38]{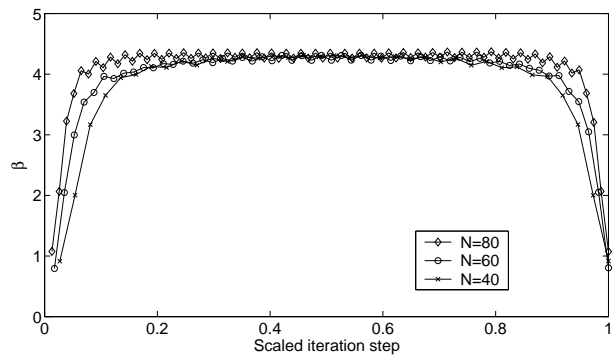}
\caption{Parameter $\beta$ as a function of rescaled iteration step using DBSS approach for $TRE_{max}=10^{-4}$, $M_{min}=16$, $U=1$, $N=40,60,80$.}  
\label{fig:qdcfig03}
\end{figure}

This standard procedure, however, does not allow an a priori control of the accuracy, while this control is possible in the DBSS approach \cite{legeza1}. Moreover, the latter
approach allows to study the relationship between the von Neumann entropy of the block and the dimension of the Hilbert space of the superblock qualitatively. Therefore,
we have repeated the DMRG calculations on the same model using the DBSS approach with $M_{min}=16$ for various values of the 
truncation error, namely for $TRE_{max}=10^{-2},10^{-3},\cdots10^{-8}$. 
In Fig.~\ref{fig:qdc_dbss_u1} we have plotted $\ln (\dim \Lambda)$, $\ln (\dim \Lambda_{SB})$ 
and the block entropy obtained after the 6th sweep for $U=1$, $N=80$ 
as a function of iteration step shifted modulo $80$.
It is clearly seen in the figure that 
due to the truncation of basis states the logarithmic corrections are cut, 
$S$ and $\ln (\dim \Lambda)$ is almost constant for a wide range of block lengths.  
A simple relationship seems to exist between the two quantities, which is analogous to  
Eq.~(\ref{eq:qdc02}). We define the shift between $\ln (\dim \Lambda)$ and $S$ as 
\begin{equation}
\beta \equiv \ln(\dim \Lambda) - S\,. 
\label{eq:qdc03}
\end{equation}
Fig.~\ref{fig:qdcfig03} shows the value of $\beta$ for $U=1$, $TRE_{max}=10^{-4}$ for 
three different chain lengths, $N=40,60,80$.  
$\beta$ is plotted as a function of the shifted iteration step rescaled by the length of the chain.
We have found that after the sixth sweep except for very short block lengths $\beta$ is practically independent of $N$   
for a given accuracy.
A similar result is expected for non-critical systems since according to Vidal {\sl et al.} \cite{vidal10} in this case the block entropy saturates
with increasing block length. 
It is worth mentioning that $\beta^\prime$ defined analogously to 
Eq.~(\ref{eq:qdc03}) but with $\ln(\dim \Lambda_{SB})$, 
\begin{equation}
\beta^\prime \equiv \ln(\dim \Lambda_{SB}) - S\,, 
\label{eq:qdc03a}
\end{equation}
is also independent of $N$ for large enough $N$.

Both $\beta$ and $\beta^\prime$ are, however, functions of $TRE_{max}$.   
This dependence has been tested    
for $U=1$ for $TRE_{max}=10^{-3}-10^{-6}$, when $\max(\ln (\dim \Lambda_{SB}))$ changed from $6\times 10^3$ to $2.1\times 10^6$. 
In this range of the truncation error $\beta$ increases proportionally to $-\ln TRE_{max}$.
Similar results have been obtained with open boundary condition but with significantly smaller block entropies and much less block states.
For larger $U$ values and for small values of $TRE_{max}$ 
the target state has often been lost and the convergence depended very much on the minimum number of  
block states $M_{min}$.  

\subsection{A new truncation procedure}

\begin{figure}
\includegraphics[scale=0.38]{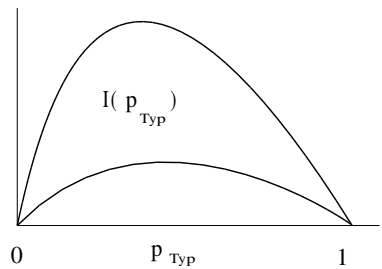}
\caption{Schematic plot of the upper and lower bonds on accessible information for a binary channel.}
\label{fig:bonds}
\end{figure}

The simple relationships given by Eqs.~(\ref{eq:qdc03}) and (\ref{eq:qdc03a}) 
seem to indicate that 
a better DMRG procedure that avoids the above mentioned problems  
should rely on the control of the von Neumann entropy of the blocks.  
In order to control the weight of retained information during the RG procedure in a more rigorous way, 
the reduced density matrix of the left or right subsystem 
is written as the sum of the density matrices of mutually orthogonal mixed states   
belonging to the typical and atypical subspaces,
\begin{equation}
\rho^{(L)} = p_{\rm typ} \rho^{(L)}_{\rm typ} + (1 - p_{\rm typ}) \rho^{(L)}_{\rm atyp}\,,
\end{equation}
where $\rho^{(L)}_{\rm typ}$ is formed from the $M$ largest
eigenvalues of $\rho^{(L)}$ and $\rho^{(L)}_{\rm atyp}$ from the remaining eigenvalues
with ${\rm Tr}\rho^{(L)}_{\rm typ} = {\rm Tr}\rho^{(L)}_{\rm atyp} = 1$.
Similar decomposition holds for the right block, and therefore in what follows the superscript $L$ and $R$ are dropped.
In usual information theory, if the message contained $\rho_{\rm typ}$ or $\rho_{\rm atyp}$ with the appropriate probabilities, the accessible 
information for such a binary channel would be less than the Kholevo bound \cite{kholevo1}
\begin{equation}
I \leq S(\rho) -  p_{\rm typ} S(\rho_{\rm typ}) - (1- p_{\rm typ}) S(\rho_{\rm atyp})\,,
\label{eq:holevo}
\end{equation}
and larger than the Jozsa-Robb-Wootters lower bound \cite{jozsa4}
\begin{equation}
I \geq Q(\rho) -  p_{\rm typ} Q(\rho_{\rm typ}) - (1- p_{\rm typ}) Q(\rho_{\rm atyp})\,.
\end{equation}
The schematic plot of the two bounds as a function of $p_{\rm typ}$ is shown in Fig.~\ref{fig:bonds}.  
It is worth to mention that other ensemble dependent bounds have been derived by Fuchs and Caves \cite{fuchs1}.  

In DMRG, however, the atypical subspace is neglected, which leads to a loss of information.
For $p_{\rm typ}$ close to unity this loss is argued to be equal to the accessible information 
which is negligible if the probability of the atypical subspace is negligible. 
This can be used to optimize the DMRG procedure.   
A modified DBSS procedure is proposed, where instead of fixing the truncation error, the Kholevo bound  
is required to be less than an $\epsilon$ fixed in advance,
and the number of states $M$ is chosen accordingly in every step.

\begin{figure}
\includegraphics[scale=0.38]{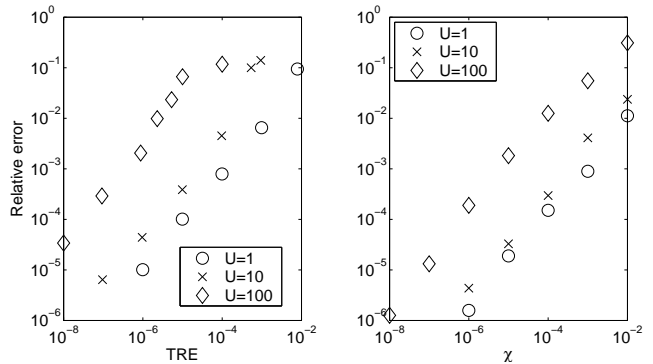}
\caption{The relative error as a function of the truncation error and the threshold value of Kholevo's bound on accessible information
obtained with the DBSS method and by Eq.~(\ref{eq:holevo}), respectively.}
\label{fig:error}
\end{figure}

We have run independent calculations for the 1-d Hubbard model for different threshold values on the upper bound 
of accessible information, denoted by $\chi$, ranging from $10^{-2}$ to $10^{-7}$ using three sweeps. 
In Fig.~\ref{fig:error} we have plotted the relative error as a function of $TRE$ and $\chi$ on a logarithmic scale for $U=1,10,100$ with $M_{min}=4$,
where the relative error is given as $(E_{\rm DMRG}-E_{\rm exact})/E_{\rm exact}$.
We have found that this selection rule is very stable even for very small values of the retained eigenvalues of $\rho_{\rm typ}$. 
Assuming that the linear relationship apparent on the right panel holds for other models as well, a  
new extrapolation procedure can be obtained. Instead of using the truncation error as proposed in Refs.~ \cite{white1,legeza0,extrap1,extrap2,extrap3}, 
the energy of the target state is extrapolated as a function of $\chi$. 


\subsection{Application to quantum chemistry DMRG}

\begin{figure}
\includegraphics[scale=0.4]{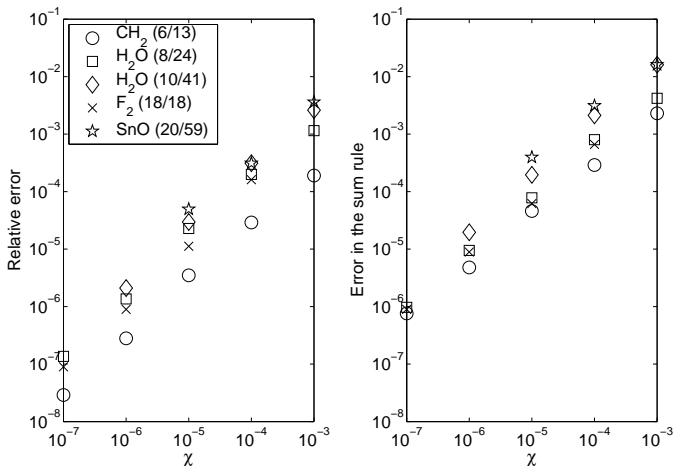}
\vskip .2 cm
\caption{The first panel shows the relative error as a function of the threshold value of Kholevo's bound on accessible information
obtained by Eq.~(\ref{eq:holevo}). The second panel shows 
the error in the sum rule given by Eq.~(\ref{eq:sumrule2}).}
\label{fig:error_qc}
\end{figure}

We have tested the new truncation procedure  
presented above by  performing DMRG calculations
on various molecules up to 59 lattice sites. We have 
used the dynamically extended active space procedure (DEAS) \cite{legeza3} 
and the DBBS approach by controlling the change of von Neumann entropy
during the truncation procedure at each renormalization step.  
In all calculations $M_{min}=4$ was used and during the first half sweep the right block was represented 
by $256 - 512$ states. The maximum number of block states that our program could treat is in the range of 
2000-3500, strongly depending on the system size.
Our results on various molecules are shown in Fig.~\ref{fig:error_qc}. 
We have found again that on a logarithmic scale the relative error is a linear function of the accessible information
even for large threshold values.
 
For very small system sizes, like CH$_2$ (6/13) the number of block states has been overestimated by using the upper bound on the
accessible information. Therefore, we have also tried to control the lower bound on accessible information.
For large threshold values, the lower bond on accessible information has also provided similar results but
with less number of block states. For smaller threshold values, however,  
we have found it to be very unstable for increasing $M$ with larger values of $p_{\rm typ}$.

When molecules are studied in quantum chemistry application,
``lattice sites'' carry different quantum information as   
described by a non-constant value of the site entropy function. 
This means that the entropy profile
of the bi-partite partitioning of the finite system depends very much on the ordering of the lattice sites.
Since the molecules are non-critical models, one can obtain the entropy profile of the 
exact solution of the finite system with an exponentially small error using a limited number of block states.
Since the block entropy is related to the size of the superblock Hilbert space, 
in order to increase the  
efficiency of the DMRG method one has to reduce the block entropy profile of the exact solution by optimizing ordering.
For larger systems we have found that the optimization procedure outlined in Ref.~\cite{legeza3} 
based on the site entropy profile alone, does not lead to optimal ordering
and it often blocked the convergence of the DMRG method. 
This problem related to the competition between entanglement localization and 
interaction localization will be investigated in a subsequent paper.

\section{Relationship to quantum information generation}

In the DMRG procedure, during the renormalization step the block $B_L$ is formed of the
subblock $B_l$ and the $l+1^{\rm th}$ site. Denoting by $S_L(l)$ the entropy of the
left subblock of length $l$ and by $S_{l+1}$ the entropy of the $l+1^{\rm th}$ site,
the change of the block entropy by forming a larger block, $B_L(l+1)$, is given as
\begin{equation}
S_L(l) + S_{l+1} + I_L(l)  = S_{L}(l+1)\,,
\label{eq:information2}
\end{equation}
where the so-called mutual information $I_L(l)$ quantifies the correlation between the subsystem and the site. 
A similar relation holds for the right block, given as
\begin{equation}
S_R(r) + S_{l+2} + I_R(r)  =  S_{R}(r+1)\,.
\label{eq:information3}
\end{equation}
The unitary operation applied on the basis states of the $B_l$ $\bullet$ composite system is a type of LOCC 
(local quantum operations or classical communications)
operation, i.e., it cannot increase the entanglement between $B_L$ and $B_R$ blocks which has also been related
to entanglement distillation protocols \cite{delgado}.
Therefore, the entropy of the $B_l$ $\bullet$ composite system remains unchanged or  
decreased by forming the larger block $B_L$. 
The quantum information generated by the renormalization procedure of the forward sweep can be measured using Eq.~(\ref{eq:information2})  as 
\begin{equation}
I_L(l) = S_L({l+1}) - S_L(l) - S_{l+1}\,,
\end{equation}
where $l$ runs from $1$ to $N-1$. 
Analogously the information generated by the backward sweep can be derived using Eq.~(\ref{eq:information3}). 

If an effective system of length $N+2$ is formed by adding  
two non-interacting sites to the right end of the chain,  
all blocks containing 1 to $N$ lattice sites of the original system can be formed by the forward sweep. 
The total information gain of a full half sweep can be calculated as $\sum_{l=1}^{N-1} I_L(l)$. 
The same holds for the backward sweep as well when the two non-interacting 
sites are attached to the left end of the chain. 
It is easy to show that if all $q^l$ and $q^r$ basis states
of the blocks are kept at each iteration step, i.e., no truncation is applied,   
a sum rule holds, which relates
the total information gain within a full half sweep and the sum of site entropies given as
\begin{equation}
\sum_{l=1}^{N-1} I_L(l) = - \sum_{l=1}^N S_l \,,
\label{eq:i_tot}
\end{equation}
where we have used $S_L(1)=S_1$ and $S_L(N)=0$.
This equality, however, does not hold in practical DMRG calculations. First the blocks contain only a limited subset of
the block states thus for a given site $S_l$ changes for each half sweep as the method converges to the attractor. In addition, during the
renormalization process $S_L({l+1})$ is reduced to $S_L^{\rm Trunc}(l+1)$  due to the truncation of basis states, thus 
$I_L(l)$ is also a function of subsequent sweeps.
However, once the DMRG method has converged, i.e., 
subsequent DMRG sweeps do not change $S_L(l)$ and $S_l$, the following equality should hold to a good accuracy
\begin{equation}
\sum_{l=1}^{N-1} I_L(l) \simeq - \sum_{l=1}^N S_l + \sum_{l=2}^{N} \left( S_L({l}) - S_L^{\rm Trunc}(l) \right) \,.
\label{eq:i_tot_trunc}
\end{equation}
An analogous relationship holds for the backward sweep as well.
 
As one sees in Fig.~\ref{fig:bonds}  
the decrease of the loss of information goes together with the decrease of the contribution of the atypical states. 
Although the actual shape of the upper and lower bounds depends on $U$, we have found that when only the first few eigenvalues of $\rho$  
are retained, $p_{\rm typ}$ is already very close to unity.   
Therefore, the loss of information could be
equally controlled
by requiring that $\chi \equiv S(\rho)-S(\rho_{\rm typ})$ should become less than $\epsilon$ in which case Eq.~(\ref{eq:i_tot_trunc}) leads to
\begin{equation}
\sum_{l=1}^{N-1} I_L(l) + \sum_{l=1}^N S_l <  (N-1)\epsilon.
\label{eq:sumrule2}
\end{equation}
Eq.~(\ref{eq:sumrule2}) can be used as an alternative procedure to check the convergence of the DMRG method.  

Numerically we have found a very similar result to the one shown in the previous section, that the relative error and 
$\sum_l I(l) + \sum_l S_l$ are linear functions of $\chi$ on a logarithmic scale. 
The second panel of Fig.~\ref{fig:error_qc} shows the error in the sum rule given by Eq.~(\ref{eq:sumrule2}) 
for the calculation presented in Sec.~III D. 
This has been 
found to be one order of magnitude worse than the discarded weight of the von Neumann entropy, as expected.

\section{Summary}

We have studied 
the density-matrix renormalization group method from the point of view of quantum data compression.
The method has been applied to the half-filled 1-d Hubbard model in real space and to 
various molecules. 
Our findings are listed below:

(1) We have shown an explicite relationship between the dimension of the Hilbert space of DMRG blocks and block entropy 
for any fixed accuracy or alternatively between the superblock Hilbert space and block entropy.

(2) We have 
presented a more rigorous truncation procedure based on Kholevo's theory on accessible information. 
This method also provides a new extrapolation method. We have found numerically that 
on a logarithmic scale the relative error is a linear function of the threshold value on
the accessible information.  

(3) We have also studied models when lattice site entropies are not equivalent  
and demonstrated the efficiency of our new truncation procedure and criteria of convergence.

(4) We have presented a sum rule which relates the sum of site entropies to the total information generated
within a half sweep of the DMRG method. It has been shown that this could be used as an alternative test
for convergence.

The application of other concepts of quantum information theory \cite{delgado1} to DMRG might also prove to be useful when more complicated systems are studied.
A more rigorous study of entanglement distillation and purification protocols for systems when site entropies are not equivalent
could lead to further optimization of the DMRG method. 
Another natural extension of the present work would be to use non-orthogonal states, however, the DMRG implementation of such a 
problem is rather complicated \cite{nonort}.

\acknowledgments

This research was supported in part by    
the Hungarian Research Fund(OTKA) Grants No.\ 32231, 43330 and 46356. 
The authors thank M. Reiher and J. R\"oder for providing the integral matrices and the ground-state energy of the SnO and H$_2$O molecules.
\"O.~L.~acknowledges
useful discussions with F.~Gebhard, R.~Noack and G.~F\'ath and computational support from the Dynaflex Ltd.  


\end{document}